%% file: ms.tex
\documentclass[conference]{IEEEtran}
\IEEEoverridecommandlockouts
\usepackage{cite}
\usepackage{amsmath,amssymb,amsfonts}
\usepackage{algorithmic}
\usepackage{graphicx}
\usepackage{textcomp}
\usepackage{verbatim}
\usepackage{graphicx}
\usepackage{subfig}
\usepackage{booktabs} 
\usepackage[ruled]{algorithm2e} 
\usepackage{xspace}
\usepackage{fixltx2e}
\usepackage{xcolor}
\usepackage{paralist}
\usepackage{amsmath}
\usepackage{hyperref}
\usepackage{slashbox}

\input{commands}
\def\BibTeX{{\rm B\kern-.05em{\sc i\kern-.025em b}\kern-.08em
    T\kern-.1667em\lower.7ex\hbox{E}\kern-.125emX}}
\begin{document}

\title{\paperTitle{}
}

\author{\IEEEauthorblockN{Vassilis Vassiliadis}
\IEEEauthorblockA{vasiliad@uth.gr}
\and
\IEEEauthorblockN{Konstantinos Parasyris}
\IEEEauthorblockA{koparasy@uth.gr}
\and
\IEEEauthorblockN{Christos D. Antonopoulos}
\IEEEauthorblockA{cda@uth.gr}
\and
\IEEEauthorblockN{Spyros Lalis}
\IEEEauthorblockA{lalis@uth.gr}
\and
\IEEEauthorblockN{Nikolaos Bellas}
\IEEEauthorblockA{nbellas@uth.gr}
\and
\IEEEauthorblockN{}
\and
\IEEEauthorblockN{
\IEEEauthorblockA{
\textit{Centre for Research and Technology Hellas} \\
Volos, Greece}
}
\and
\IEEEauthorblockN{
\IEEEauthorblockA{
\textit{Department of Electrical and Computer Engineering} \\
\textit{University of Thessaly}\\
Volos, Greece}
}
}

\maketitle

\input{abstract.tex}


\input{1_introduction.tex}

\input{2_unreliable.tex}

\input{3_methodology.tex}

\input{4_unreliable_modeling.tex}

\input{5_metric.tex}

\input{6_results.tex}

\input{7_related_work.tex}
\input{8_conclusions.tex}

\bibliographystyle{IEEEtran}
\bibliography{references}

\end{document}

%% file: commands.tex
\newcommand\metric[1]{EEOP{#1}\xspace}

\newcommand\accuracyIncorrect{{\emph{Detection~Coverage}\xspace}}

\newcommand\errorAccepted{{\emph{$\overline{MRE}$}\xspace}}

\newcommand\headerAccuracy[1]{&Rejection&Acceptance&Rejected&Accepted\\
Name&Accuracy&Accuracy&Error (\%)&Error (\%)\\}

\usepackage{array}
\newcolumntype{L}[1]{>{\raggedright\let\newline\\\arraybackslash\hspace{0pt}}m{#1}}
\newcolumntype{C}[1]{>{\centering\let\newline\\\arraybackslash\hspace{0pt}}m{#1}}
\newcolumntype{R}[1]{>{\raggedleft\let\newline\\\arraybackslash\hspace{0pt}}m{#1}}

\newcommand\paperTitle{Artificial neural networks for online error detection\xspace{}}


%% file: abstract.tex
\begin{abstract}
Hardware reliability is adversely affected by the downscaling of semiconductor devices and the scale-out of systems necessitated by modern applications. Apart from crashes, this unreliability often manifests as silent data corruptions (SDCs), affecting application output. Therefore, we need low-cost and low-human-effort solutions to reduce the incidence rate and the effects of SDCs on the quality of application outputs. We propose Artificial Neural Networks (ANNs) as an effective mechanism for online error detection. We train ANNs using software fault injection. We find that the average overhead of our approach, followed by a costly error correction by re-execution, is $6.45\%$ in terms of CPU cycles. We also report that ANNs discover $94.85\%$ of faults thereby resulting in minimal output quality degradation. To validate our approach we overclock ARM Cortex A53 CPUs, execute benchmarks on them and record the program outputs. ANNs prove to be an efficient error detection mechanism, better than a state of the art approximate error detection mechanism (Topaz), both in terms of performance ($12.81\%$ CPU overhead) and quality of application output ($94.11\%$ detection coverage).
\end{abstract}


%% file: 1_introduction.tex
\section{Introduction}
\label{sect:1_introduction} 

The ongoing down-scaling of transistors as well as the reduced operating voltages combined with scaling-out systems led to a dramatic rise in susceptibility of processors~\cite{Borkar:2005:DRS:1108266.1108285, 5755188} and systems~\cite{DARPA1,amarasinghe2009exascale} to faults. As computing systems are now ubiquitous, a system malfunction is likely to have more severe financial and social impacts than ever before. In one incident, a single soft error crashed a system farm, while in another case a single soft error disrupted the operation of a billion dollar automotive factory \cite{ziegler}. Several other incidents have been reported \cite{Mukherjee:2003:SMC:956417.956570,normand1996single}. Crashes are the obvious side effects of hardware faults. Silent Data Corruptions (SDCs) may be even more dangerous because they are not immediately observable. Unfortunately, SDCs are also expected to appear more frequently ~\cite{cappello2009toward}. 

Increasing the resilience of computing systems has become a primary concern. Hardware architects have introduced several error protection mechanisms and guardbands. However, those methods are not sufficient~\cite{Narayanan:2010:SSP:1870926.1871008}. For example, conventional circuit design techniques for high performance tune all timing paths to the critical path. Consequently, when a single path fails more are prone to failure, thereby negatively affecting the overall reliability of the hardware~\cite{patel2008cmos,sartori2009alleviating}.  Moreover, improving the reliability of computations by guardbanding and redundancy not only increases the cost of the manufacturing process, but also constraints the performance of the designs, because they are pessimistically configured for the worst-case scenario (critical path).

A major challenge is detecting quality-degrading errors before they irreversibly modify application state. Ideally, a detection mechanism should be able to distinguish between correct and incorrect computation results, at a very low performance overhead.  Fortunately, lower levels of the system such as circuit, micro-architecture and architecture largely mask hardware faults before they incur changes to the application layer~\cite{Wang:2004:CET:1009382.1009722}. Out of those which manifest to the application state, it is important to detect the ones which significantly affect the output quality of the application, and correct them before they propagate to the output. Depending on application characteristics, minor deviations from correct intermediate results may not be worth paying the correction cost. There is a large body of work on approximate computing that gracefully trades-off computation accuracy for improved performance, energy- and power-efficiency~\cite{Topaz2015, Chisel2014, Rinard_ICS2006, CGO2016,Ringenburg:2015:MDQ:2694344.2694365, Khudia:2015:ROQ:2872887.2750371, McAfee:2015:EFG:2738600.2738616, Esmaeilzadeh:2012:NAG:2457472.2457519}.

We introduce a methodology for automatic error detection, based on Artificial Neural Networks (ANNs).  
ANNs have been successfully deployed in pattern matching and classification applications, sometimes even outperforming human accuracy~\cite{he2015delving, assael2016lipnet}. Typically, ANNs are treated as black boxes which, given enough observational data, 
can become very efficient at classifying data and approximating functions. 
Given the configurability of their architecture, one can flexibly trade-off performance and classification accuracy by modifying the number of layers and the number of nodes per layer. 

This paper contributes the following: 
\begin{inparaenum}[(i)]
\item We use ANNs to detect hardware errors during program execution. Furthermore, we compare our approach with Topaz, a state of the art approximate error detection mechanism~\cite{Topaz2015}.
\item We evaluate the overhead of our approach and its impact on the quality/accuracy of the end result, on a set of $6$ applications from the domains of imaging, finance and physics, via software fault injection experiments. We expand our case-study to simulate the use of overclocked and unreliable CPU cores as a means to optimize the performance of applications.
\item We present a case study on ANN-based error-detection for $4$ applications that execute under real unreliable conditions on an ARM Cortex A53 CPU.
\end{inparaenum}

We explore both software fault injection as well as real hardware that operates under unreliable conditions. On the one hand, real hardware, configured to operate outside its normal working envelope, leads to the manifestation of real errors. Unfortunately, this requires support from all levels of the system stack, from the application level all the way down to the hardware level. Additionally, errors seldomly appear and thus it is difficult to investigate the use of error detectors. As such, it is difficult to study the behaviour of long running applications on real unreliable hardware. Therefore, we first investigate the efficacy of our method for large input data sets under simulated unreliable conditions. Then, we evaluate the use of the very same error-detectors for applications which execute using real unreliably configured hardware but with smaller input data sets.

We show that ANNs can act as efficient error detectors, offering a good trade-off between accuracy and execution overhead. Also, our ANN generation method is semi-automatic and requires minimum human effort.

The rest of the paper is structured as follows.
Section~\ref{sect:2_concept} introduces the conceptual model for the computations we target in this work. Section~\ref{sect:3_methodology} details our approach to automatic error detection using ANNs. Section~\ref{sect:4_unreliable_modeling} describes modeling of unreliable execution and Section~\ref{sect:5_metric} introduces an error detector fitness metric. In Section~\ref{sect:6_results} we evaluate our methodology. Section~\ref{sect:7_related_work} provides an overview of related work. Finally, Section~\ref{sect:8_conclusions} concludes the paper.

%% file: 2_unreliable.tex
\section{Conceptual model}
\label{sect:2_concept}

For the purpose of this work we assume a task-based programming model, where the computation is broken down to smaller tasks. The inputs and outputs of each task are explicitly defined. Tasks can be executed in parallel, subject to their data dependencies. 

Some tasks are executed on potentially unreliable hardware. Consequently, a task may experience an unexpected runtime error and crash, enter an endless loop, or manage to complete but produce incorrect output. The first two types of failures can be detected by the runtime system through the standard operating system mechanisms and watchdog timers, respectively. On the contrary, the last type of failure, also referred to as a silent data corruption (SDC), is typically detected via error-detection code or some form of voting mechanism such as Triple Redundancy Modular check (TRM)~\cite{lyons1962use}. The former uses code (typically supplied by the application programmer) that inspects the output of the task to infer whether it is correct or not. The latter involves multiple executions of the same task on different execution vehicles and subsequent comparison of the respective results. Implementing result checking can become tedious, whereas TRM may introduce extreme overheads.

We assume that an error detector executes on reliable hardware and never experiences SDCs. However, the error detector is not guaranteed to always make correct decisions. This affects both the result quality/accuracy and the execution time of applications. On the one hand, if an incorrect task output is accepted as correct, it may eventually corrupt the global state of the application, which will degrade, potentially significantly, the quality/accuracy of the end result. On the other hand, correct output may be classified as incorrect. This will unnecessarily trigger the correction/repair mechanism, increasing the application execution time of the application. If correction is performed via approximation unnecessary corrections may even degrade the output quality. 

Designing and implementing a fully accurate (perfect) error detector is too costly to be practical in the general case, or even impossible for tasks that perform non-deterministic computations. Employing a perfect error detector may also be an overly conservative (and expensive) choice for computations that do not need to produce fully accurate results. In this case, even inaccurate error detectors can lead to acceptable results without penalizing performance. This is where the usage of ANNs is particularly promising. Our goal is to exploit the observation that most computations will finish successfully without any observable error~\cite{Parasyris:2017:SPE:3086564.3058980} and try to detect and correct the most quality-degrading errors. This is the reason why we opt not to approximate code.

Error correction is not the focus of this paper. For simplicity, we assume that task outputs are not propagated/committed to the global state of the program before error checking completes. Also if the output of a task is considered to be incorrect then its output is ignored, and the task is re-executed in a reliable configuration. While this incurs a high error recovery cost, it can be acceptable as long as the error detector is successful in detecting only those errors that would have a big impact on the end result of the computation. An alternative error-correction approach is to employ an approximate version of the task code in order to reduce the execution cost at the price of producing inaccurate task output~\cite{CGO2016, McAfee:2015:EFG:2738600.2738616, Esmaeilzadeh:2012:NAG:2457472.2457519}. Since this would also degrade the end quality/accuracy of the entire computation, it would be harder to evaluate the impact of the error detection mechanism, which is the focus of this paper. Therefore, in the context of this work we correct errors through accurate and reliable task re-execution.

%% file: 3_methodology.tex
\section{Artificial neural networks for automatic error detection}
\label{sect:3_methodology}

\begin{figure*}[t]
\centering
\includegraphics[width=0.95\textwidth]{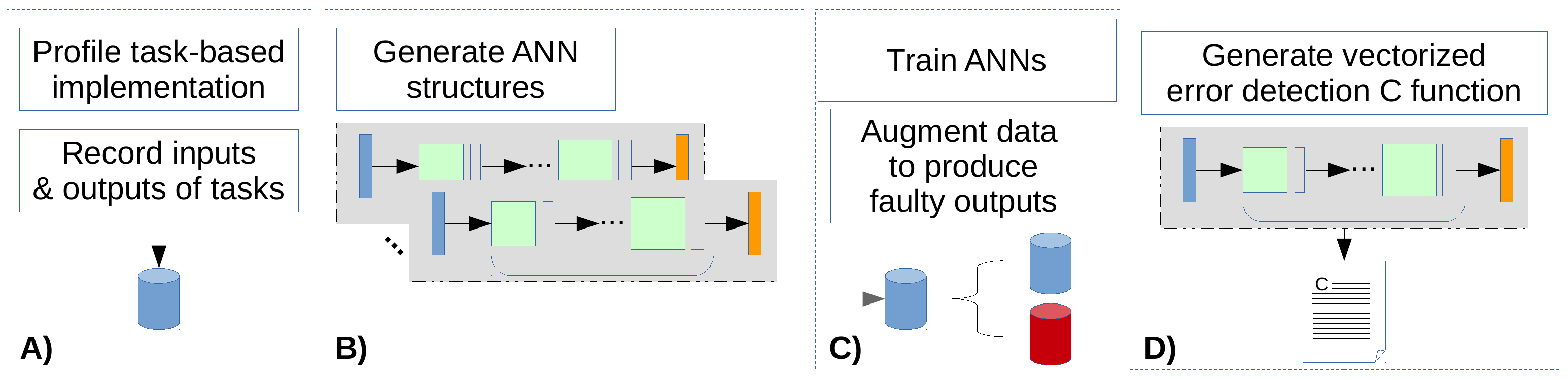}
\caption{Stages of proposed automatic error detection methodology.}
\label{fig:methodology_flow}
\end{figure*}

The manual implementation of accurate and low cost error detection is a time consuming and intricate process. The developer should be highly familiar with the application in order to be able to take educated decisions on how to detect errors in the output of each task. Also, it can be quite hard to find the desired balance between execution complexity/cost and error detection accuracy. 

Previous work has mainly explored the use of manually implemented, low-cost, but potentially approximate  error-detectors~\cite{Topaz2015, hari2012low,kadric2014energy,grigorian2014dynamically}, and methods which try to detect all possible SDCs, even if they do not necessarily affect the application output quality~\cite{feng2010shoestring, lu2014sdctune, didehban2016nzdc, laguna2016ipas}. In contrast, we propose to use ANNs as inaccurate error detectors. ANNs require little manual intervention, offer the opportunity to trade-off error detection accuracy with performance overhead in a flexible way and are widely used for classification purposes.

Typically, solving a problem through the use of ANNs involves training multiple ones and selecting the fittest.
Our methodology is illustrated in \textit{Figure~\ref{fig:methodology_flow}}, it consists of four main steps: a) gathering data for the training process, b) deciding the architecture of the ANNs, c) training the ANNs, and finally d) deploying the ANNs. This section describes this process, for constructing ANNs that will be used as error-detectors.

\subsection{Collecting training data}
We begin by partitioning the application code into fine-grained tasks. The outputs of those tasks are subject to error detection. 
The rationale is that if the task size is rather small, then its output is more likely to exhibit a detectable pattern. We train ANNs to identify this pattern (or deviations from it).

The data which an ANN processes to determine whether the output of a task is correct or incorrect is called as \emph{feature vector}. The feature vector always includes the task output. This can be sufficient for tasks with distinctive output patterns, irrespective of their input. If the output pattern is highly input-sensitive, the feature vector also has to include parts of the task input, or in the extreme case the entire input. In the latter case the ANN estimates whether the provided task output is correct for the given the task input. This is an easier problem than the problem of learning to approximate the function which the original task performs because, in our case, the ANN detector does not necessarily need to predict the exact output. In this work, the application developer manually defines the feature vector, although in principle this step could also be automated through some dimensionality reduction method, for example principal component analysis.

The application is then executed reliably for different inputs. In every execution, we record the feature vector of each task. This data is aggregated for each task across all executions to produce a so-called \emph{profile} data set. We select $\sim 90\%$ of the profile data set to construct the \emph{training} set, while the remaining $\sim 10\%$ of the profile set is used as the \emph{test} set. We update the weights of the ANNs using the \emph{training} set and perform early stopping by monitoring the loss on the \emph{test} set (Section~\ref{sect:3_training}).

\subsection{ANN structure}
\label{sect:3_annstruct}
For online error detection, we consider the least complicated type of ANNs: Multilayer Perceptrons that consist of InnerProduct (IP) layers. Rectified Linear Units (ReLUs) serve as activation functions between IP layers. 

IP layers accept $M$ inputs and produce $N$ outputs. Each such layer contains an $MxN$ matrix of weights ($W$) and a vector ($\vec{b}$) that contains $N$ bias values. Assuming that the vector $\vec{x}$ is input to an IP layer, the resulting output vector is $\vec{y}=W*\vec{x}+\vec{b}$. ReLU activation layers are placed between two IP layers to introduce non-linearities which improve the generalization of the ANN as well as classification performance. A ReLU with an input vector $\vec{x}$ and output vector $\vec{y}$ computes $y_i=max(0, x_i) \; \forall \; x_i \in \vec{x}$.

\begin{figure}[tb]
\centering
\includegraphics[width=\linewidth]{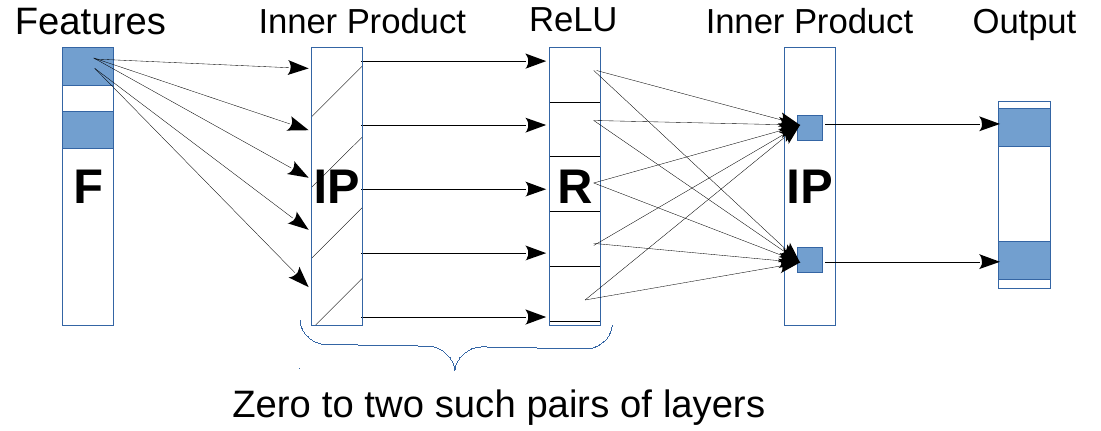}
\caption{Structural template for the generated ANNs. The input is the feature vector of the task. The output consists of two values corresponding to the one-hot classification of the task output as correct or incorrect.
} \label{fig:ann_template}
\end{figure}

\begin{table*}
\begin{center}
\begin{tabular}{ |l|l| } 
\hline
\textbf{Notation} & \textbf{Structure} \\
\hline
N,2 & $f[N] \rightarrow IP[2] \rightarrow out[2]$ \\
N,B/2,2 & $f[N] \rightarrow IP[B/2] \rightarrow ReLU \rightarrow IP[2] \rightarrow out[2]$ \\
N,B,2 & $f[N] \rightarrow IP[B] \rightarrow ReLU \rightarrow IP[2] \rightarrow out[2]$ \\
N,B*2,2 & $f[N] \rightarrow IP[B*2] \rightarrow ReLU \rightarrow IP[2] \rightarrow out[2]$ \\
N,B/2,B/2,2 & $f[N] \rightarrow IP[B/2] \rightarrow ReLU \rightarrow IP[B/2] \rightarrow ReLU \rightarrow IP[2] \rightarrow out[2]$ \\
N,B,B/2,2 & $f[N] \rightarrow IP[B] \rightarrow ReLU \rightarrow IP[B/2] \rightarrow ReLU \rightarrow IP[2] \rightarrow out[2]$ \\
N,B*2,B/2,2 & $f[N] \rightarrow IP[B*2] \rightarrow ReLU \rightarrow IP[B/2] \rightarrow ReLU \rightarrow IP[2] \rightarrow out[2]$ \\
\hline
\end{tabular}
\caption{The seven different ANNs used for error detection. $f$ is the feature vector of the task, $N$ is the size of the feature vector, $B$ is the power of 2 closest to $N$, and $out$ is the 2-dimensional output. The dimension of the input/output vectors and IP layers are given in brackets}
\label{tab:ann_struct}
\end{center}
\end{table*}

For every task function for which we wish to build an error detector, we experiment with various ANN structures, according to the template shown in {\it Figure~\ref{fig:ann_template}}. The input of the first IP layer is always the feature vector for the task in question (the size of the vector depends on the task). The last IP layer produces exactly two values, to encode the confidence of the ANN that a task output is correct or not. Our ANNs are \emph{one-hot classifiers} where each component of the output vector gives a score for the correctness and the incorrectness of the input feature vector, respectively. The component with the highest score is considered to be the output of the ANN classifier. In the unlikely event of a tie, we assume that the output is incorrect.
With each ANN we explore a different combination for: (a) the number of IP layers and ReLUs, and (b) for the sizes of these layers, for which we try out different powers of 2 as this facilitates vectorization. These ANNs vary in their internal computational complexity and may also have different error-detection accuracy. More specifically, for each task we consider the 7 ANNs structures listed in {\it Table~\ref{tab:ann_struct}}. 
These ANN structures are produced automatically, based on the size of the feature vector. 

\subsection{Training the ANNs}
\label{sect:3_training}
We train the ANNs using Caffe~\cite{jia2014caffe}. Training is done in \emph{epochs}: an epoch is over when the training set has been fed-forward, the loss function for the resulting ANN outputs has been evaluated, and the weights have been adjusted by back-propagation of the loss on the test data. The completion of the training process is decided using the following heuristic. Initially, when training starts, we issue 100 \emph{tickets}. At the end of each epoch we check the loss of the ANN on the \emph{test} data. If the test loss has decreased compared to the previous epoch, then the number of tickets is increased by $1$ (to a maximum of 100), else it is decreased by $2$. When there are no tickets left, the ANN is assumed to have reached an acceptably low loss, and training terminates. We reward the reduction of loss to assist the recovery from local test-loss minima. After all, consecutive training epochs may increase the test loss before it is eventually reduced beyond its past minimal value.

For a task output that comprises $N$ values, each one consisting of $k$-bits, the possible different incorrect output variants are $N*(2^{k}-1)$. If during training one considers all possible incorrect values for a single correct output, this would bias the ANN to always infer that a feature vector is incorrect, as that would minimize the loss function. One the one hand, it is possible to counter-balance all the possible incorrect output patterns by scaling the loss for the correct feature vectors. Unfortunately, taking into account all possible errors would dramatically increase the size of the training and test data sets and lead to unacceptably long training times.

As a more practical approach, we rely on data augmentation
to periodically generate data sets containing equal numbers of \emph{correct} and \emph{incorrect} feature vectors. 
We perturb the correct feature vectors (from the profile data) to produce incorrect ones, every few training epochs. This way the ANNs are trained to classify feature vectors that have a strong similarity (but are not necessarily identical) to the correct ones as \emph{correct}, and to classify widely different patterns as \emph{incorrect}.

\subsection{Deployment}

We have implemented a Caffe-to-C python script which takes an ANN model from Caffe and produces C code that performs just the inference (feed-forward) operation. The C code consists mostly of {\it gcc} vector extension intrinsics, and we rely on {\it gcc} to automatically generate vectorized implementations for maximum performance. In terms of operations, IP layers comprise floating point additions, subtractions and multiplications. ReLUs can be implemented via the gcc built-in function $fmaxf()$, or using binary arithmetic. 

Finally, we add the code of the ANN-based error-detectors in the application. These are invoked post task completion, to check their output and decide whether they need to be re-executed. However, recall that tasks are intentionally fine-grained, and it is unrealistic for the application to be executed at such an extremely fine level of granularity. Due to the system-level task management overheads, this would incur significant performance penalty. To reduce this overhead, we group a large number of independent tasks into large \emph{gangs}, which are the actual, and much coarser, scheduling unit for the underlying runtime system. When a gang completes its execution, the output of each task is checked individually via the corresponding ANN, and if it is classified as incorrect then the task is flagged for re-execution. 

We also explore \emph{batched} error-detection by checking for errors in the aggregated outputs of several tasks. A task batch typically comprises just a few tasks (which can have dependencies), and for all practical purposes it can be treated as if it was a single, coarser task. The process for producing the respective error detection ANNs remains similar to the one described above. Application profiling and ANN training are exactly the same. The difference is that the feature vector has to be defined to include the aggregated outputs of the task batch and possibly parts of the aggregated inputs.

%% file: 4_unreliable_modeling.tex
\section{Modeling unreliable execution}
\label{sect:4_unreliable_modeling}
We present 3 case studies: 2 fault injection campaigns on the Intel i7 4820k CPU, and a set of unreliable execution experiments on the ARM cortex A53 CPU.

\subsection{Software fault injection}
\label{sect:fault_injection}
To perform software-fault injection we use the methodology described in~\cite{Parasyris:2017:SPE:3086564.3058980}. A fault-injection experiment injects random multiple-bit-flip faults on architecturally visible registers once every $10^7$ cycles. 
For the software fault injection campaigns we perform $192000$ experiments, for a confidence level of $99\%$ and average margin of error $1.025\%$ for each of the $6$ benchmarks. The runtime system executes the main application thread under reliable conditions and schedules tasks on 4 worker threads which are mapped on the cores of the Intel i7 4820k CPU. 

\subsection{Real unreliable execution}
\label{subsect:unreliable_execution}
For our third case study we use the ARM cortex A53 CPU of the Raspberry PI model 3b which we configure to execute unreliably. During execution time the CPU receives input for a gang-of-tasks, which it processes unreliably, and finally it returns the task outputs to a master PC that records the inputs and outputs of the tasks.

We use these outputs of true unreliable tasks but we perform error detection \& correction as well as performance measurements using the Intel i7 CPU. This enables us to detect the presence of real errors while using a consistent performance model. For each benchmark we collect $1000$ sets of task outputs which contain more than 1 incorrect task output.

%% file: 5_metric.tex
\section{Error detection fitness metric}
\label{sect:5_metric}
To evaluate our error detection method we consider the efficiency of the classifier to detect errors accurately as well as the overall overhead introduced by error detection and correction. The performance overhead of the ANN error detection mechanism depends on both its computational complexity and its level of precision and recall. For every \emph{incorrect} task, the cost of performing corrective action (task re-execution in our evaluation) is added on top of the detection overhead itself.

Table~\ref{tab:metricsTable} presents the metrics that we used to evaluate our framework. 
\metric{} can be used to choose the fittest out of a set of error detectors. The lower the \metric{} value the better the detector is.

\begin{table}
\footnotesize
\begin{tabular}{ | L{2cm} || L{4.8cm}  | }
\hline
\textbf{Metrics} & \textbf{Description}\\ \hline
\textbf{TPR}: True Positive Rate  & The detector correctly considers the task output as incorrect. (The output is indeed incorrect) \\ \hline 
\textbf{FPR}: False Positive Rate & The detector falsely considers the task output as incorrect (the output is correct)\\ \hline
\textbf{TNR}: True Negative Rate & The detector correctly considers the task output as correct\\ \hline
\textbf{FNR}: False Negative Rate  & The detector falsely considers the output as correct (The output is incorrect)\\ \hline  
\textbf{MRE}: Missed Relative Error& The average relative error across all \textit{FNR} outputs.  \\ \hline  
\textbf{EE}: Expected Error & Quantifies the average relative error of \textit{FNR} classifications  $EE = FNR*MRE=(1-TPR)*MRE$   \\ \hline  
\textbf{Overhead:} Error Detection and Correction Overhead & The percentage of cycles spent to detect and correct errors with respect to the cycles required to execute a benchmark under reliable conditions.\\ \hline  
\textbf{EEOP}: Expected Error/Overhead Product& Combines the accuracy and the overhead of the detector  $EEOP = EE * Overhead$   \\ \hline  
\end{tabular}
\caption{Metrics used to evaluate our methodology}
\label{tab:metricsTable}
\end{table}

Note that the \metric{} metric, as defined in Table~\ref{tab:metricsTable}, may produce acceptable scores even for error-detectors that have very large overheads. Such overheads might occur due to either the complexity of the error-checking mechanism or an abnormally high False Positive Rate (FPR). 
We modify \metric{} so that it discards inefficient error-detectors using a user-supplied value ($\epsilon$)\footnote{In our experiments, we set $\epsilon$ equal to $33\%$} that specifies the highest tolerated error detection and correction Overhead:\begin{equation}
\small
\textrm{\it \metric{}}  = \begin{cases}\textrm{\it Expected Error}*Overhead, & Overhead \leq \epsilon \\
\infty, & Overhead > \epsilon
\end{cases}
\label{eq:error_inefficiency_stage2}
\end{equation}

\metric{} is a composite metric, that does not directly translate to a physical property.  Reasoning on individual metric scores is invalid. However, the comparison of \metric{} values for different error detectors enables automatic systems to evaluate them using a single metric that combines both overhead and error-detection related properties. This is particularly useful in our case, because we automatically generate a number of ANN error detectors which typically represent different trade-off points between output quality and execution overhead. \metric{} enables us to automatically select the fittest error detector out of many, without requiring human intervention beyond specifying the maximum overhead threshold ($\epsilon$).

%% file: 6_results.tex
\section{Evaluation}
\label{sect:6_results}

In this section, we discuss the experimental evaluation of the ANN error detection methodology and we compare our methodology against Topaz~\cite{Topaz2015}, both in terms of accuracy and performance. Topaz executes computations using a heterogeneous computing platform that comprises a reliable main worker thread and multiple unreliable worker threads. It includes an approximate outlier detector which checks for the existence of errors on Abstract Output Vectors (AOVs). AOVs are constructed using either just the outputs of tasks or a developer implemented function. The latter operates on both inputs and outputs of a task to generate an AOV\footnote{In the case of ANNs, the equivalent of an AOV is a feature vector} which encodes the output vector of a task in a space of lesser dimension. Unlike ANNs, Topaz does not require an offline phase. When Topaz detects an error at the output of a task (false/true positive) it re-executes the task reliably, updates its error detection model, and then integrates the correct result in the main computation. The comparison between our offline trained error detectors and the online training approach of Topaz is particularly interesting. On the one hand, contrary to Topaz, our detectors are capable of judging whether or not a task produced outputs from the very first specimen that they encounter. On the other hand, Topaz requires a warm-up phase but it may then theoretically adapt to accommodate for shifts in the distribution on errors.

Given that detecting such shifts to the errors is not a trivial issue, Topaz updates its error detection model via a heuristic. Essentially, it forgets parts of its error detection model on a predetermined update cycle. In order to perform a fair comparison we take into account the warm-up phase of Topaz by executing large numbers of tasks for all of our benchmarks. Specifically, the validation datasets in Sections~\ref{subsect:analysis} and~\ref{subsect:poff_analysis} involve on average 5,886,250 tasks (with a median number of tasks equal to 1,846,252). In Section~\ref{subsect:pi_analysis} we use validation datasets of about 13,127 tasks (with a median of 10,000 tasks). As such, in all three case studies we treat Topaz fairly.

Both Topaz and ANNs can use more sophisticated AOVs/feature vectors which are generated via programmer provided functions that operate on the inputs and outputs of tasks. 
However, this increases the expected programmer effort.
Because the focus of this work is error detection without the involvement of the programmer, we will not explore more intricate feature vectors beyond the inputs and outputs of tasks. Furthermore, the authors of Topaz argue in favor of reducing the number of AOV dimensions via batching before performing the outlier detection test. 
We also apply batching by aggregating the AOVs/Feature-Vectors of $N$ tasks into a single batch, before checking for errors on an aggregate value produced by the outputs of tasks. If the ANN/Topaz at the batch level detect an error, all $N$ tasks within the batch are re-executed.

\subsection{Benchmarks}

\label{sect:dct_discussion}
Discrete Cosine Transformation (DCT) is used in image and video compression to transform a block of 8x8 image pixels to a block of 8x8 frequency coefficients. Low frequency coefficients are closer to the upper left corner of the 8x8 block, whereas high frequency coefficients reside in the lower right corner. A single task computes a 2x4 block of frequency coefficients. We construct the training and testing data sets using a set of images from~\cite{standard_images}. For the evaluation of both error detectors we use a third input set, the validation data set, which differs from both the training and testing data sets. For DCT, the validation data set contains images from the Image Compression~\cite{images} data set. Each task feature vector contains the $8$ DCT coefficients of the $2x4$ block as well as its offset within the $8x8$ block of coefficients. All tasks but those which compute the DCT coefficients residing in the upper left corner of the $8x8$ block are computed unreliably. The upper left corner significantly affects the final output quality. As such, it would be highly inefficient to subject the respective computations to unreliable execution conditions~\cite{CGO2016}.
A reliable execution of DCT (with quantization) has an output PSNR of $35.6916$ dB.

Blackscholes is a benchmark from the PARSEC suite \cite{bienia11benchmarking}. It implements a mathematical model for a market of derivatives, which calculates the buying and selling price of assets so as to reduce the financial risk. A task uses the Black-Scholes mathematical model to produce the price of a single asset. The training data set contains 400,000 assets and the testing-data set contains 40,000 assets. The validation data set comprises 100,000 assets. The three data sets are generated using a modified version of the PARSEC Blackscholes input generator which produces permutations of its bundled 2000 asset entries. Each data set is constructed using different data-ranges, so that the training/testing data and validation data are not the same.
In Blackscholes the feature vector consists of all $8$ inputs and output of the task.

Bonds~\cite{grauer2013accelerating} is a computational finance benchmark of the QuantLib library. In finance, a bond is an indication of indebtedness of the bond issuer to the holders. The issuer is obligated to pay the holders a debt which increases by a specified interest and/or pay the face amount at a pre-determined date referred to as the maturity date. Interest payments are deposited in intervals. Bonds includes a random bond generator used to generate 440,000 bonds for the training and testing data sets. The validation data set is 100,000 bond prices. Similarly to Blackscholes, we generate the input data using different value ranges.
The feature vector of a Bonds task comprises both its inputs and outputs.

Lulesh~\cite{LULESH2:changes} implements a solution of the Sedov blast problem for a material in 3 dimensions. It defines a discrete mesh that covers the region of interest and partitions the problem into a collection of elements where hydrodynamic equations are applied. A single task computes the hydrodynamic equations for 8 elements. The training data set profiles the execution of 4 different problem sizes\footnote{In Lulesh, the problem size determines the number of elements involved in the computation, e.g the problem size of 10 involves $10^3$ elements.} (N=5, 10, 15, and 20). The testing data set contains profile data for a problem size equal to N=18. The validation input-data is a problem size of N=50. Feature vectors for Lulesh contain the output of task, which is the computed forces for $8$ bodies, along with the time-stamp of the step which is a task input. All tasks are executed unreliably, apart from a random $10\%$ which are always executed reliably to improve numerical stability.

Inversek2j is a robotics benchmark from the AxBench suite~\cite{yazdanbakhsh2016axbench}. Inversek2j calculates the angles of a 2-joint arm using the kinematic equation. A task computes the pair of angles for a single 2-joint arm. We generate 1.1 million starting points to construct the training and testing data sets. For Inversek2j we include the  inputs and output of the task in the feature vector. 

\subsection{Case study: Real unreliable execution on ARM cortex A53}
\label{subsect:analysis_quality_degradation}

In our first case study we evaluate whether our methodology produces error detectors which generalize to real errors originating from unreliable hardware execution. To this end, we use the XM$^2$ framework~\cite{xm2_2018} which enables us to execute code under unreliable conditions. Specifically, we increase the stock CPU frequency of the ARM cortex A53 CPU which is present on Raspberry PI model 3b from 1.2GHz up to 1.4GHz to create the unreliable environment for our experiments. We then execute the tasks of each benchmark multiple times, and record the outputs of unreliable executions which differ from error-free executions. We repeat this process to collect 1000 outputs that contain errors for each different type of tasks for 4 benchmarks.

Table~\ref{tbl:pi_detailed_results} presents the results of our experiments for batched error-detection which intuitively presents itself as the worst case scenario in terms of error classification accuracy. Recall that for batched error detection the feature vector contains information for multiple tasks. We ask that the reader focuses first on the results for DCT for which we show the metrics for all of the 7 ANNs plus Topaz. The majority of the ANNs feature a high TPR and low FPR. Our detectors generalize to correctly detect the majority of real errors even though we used software fault injection during training. The remaining benchmarks have similar behaviour so in the interest of saving space we present a subset of the 7 ANNs.

Interestingly, our experiments on real unreliable execution highlight the importance of using \metric{}. For example in the case of Blackscholes it would be an understandable mistake to pick Topaz as the optimal error detector because of its extremely low FPR and high TPR. However, it will become apparent in the next following case study that Topaz actually induces a higher computational overhead in comparison with the best ANN. Clearly, it is hard to deduce the fittest error detector based on just output quality and error detection metrics.

\begin{table}[t]
\centering
\begin{tabular}{|rl|cc|c|}
\hline
\textbf{Benchmark} & \textbf{Detector} & \textbf{TPR ($\%$)} & \textbf{FPR ($\%$)} & \textbf{Quality (dB)} \\
\hline
DCT & 10, 2           & 14.43 & 14.94 & 31.97 \\
 & 10, 4, 2        & 61.59 & 3.54  & 34.26 \\
 & 10, 4, 4, 2     & 61.72 & \textbf{1.3}   & 34.60 \\
 & 10, 8, 2        & 85.95 & 3.19  & 36.56 \\
 & 10, 8, 4, 2     & \textbf{93.32} & 4.77  & \textbf{37.34} \\
 & 10, 16, 2       & 85.69 & 3.33  & 36.56 \\
 & 10, 16, 4, 2    & 92.53 & 2.62  & 35.99 \\
\hline
 & Topaz           & 53.71 & 2.64  & 37.15 \\
\hline
\textbf{Benchmark} & \textbf{Detector} & \textbf{TPR ($\%$)} & \textbf{FPR ($\%$)} & \textbf{Rel. Error ($\%$)} \\
\hline
Blackscholes & 8, 2           & 70.17 & 2.64 & 4.29 \\
 & 8, 2, 2        & 85.15 & 24.69  & 1.87 \\
 & 8, 8, 2, 2    & \textbf{88.25} &	25.29 &	\textbf{1.27} \\
\hline
 & Topaz           & 76.50 & \textbf{1.20} &	2.94 \\
\hline
\hline
Bonds & 6, 4, 2     &	28.91 &	10.60 &	0.34 \\
 & 6, 8, 2     &	27.74 &	\textbf{0.10}  &	0.32 \\
 & 6, 16, 2    &	27.60 &	0.09  &	0.33 \\
\hline
 & Topaz       &	\textbf{73.36} &	4.42&	\textbf{0.12} \\
\hline
\hline
Inversek2j & 4, 4, 2	    &   50.41  &	0.04  &	0.14  \\
 & 4, 4, 4, 2  &	5.10  &	\textbf{0.03}  &	0.27  \\
 & 4, 8, 4, 2  &	\textbf{55.26}  &	0.04  &	\textbf{0.12}  \\
\hline
 & Topaz  &	11.07  &	2.43  &	0.25  \\
\hline
\end{tabular}
\caption{Accuracy metrics, and resulting output qualities for batched-error detection. The baseline output quality for DCT is 38.05 dB, the remaining benchmarks present the relative error between unreliable executions and an error-free execution output as the baseline. Note that, True Positive Rate (TPR) indicates error coverage, and False Positive Rate (FPR) the percentage of correct tasks which were flagged for re-execution. Finally, numbers in bold indicate the best in their category (per-benchmark)}
\label{tbl:pi_detailed_results}
\end{table}

\subsection{Case study: Performance and quality evaluation}
\label{subsect:analysis}
We use $6$ benchmarks from the domains of imaging, finance, and physics. For all benchmarks but two (DCT and Sobel), the quality metric is the Relative Error between the output of the unreliable execution and the error-free execution. In DCT we measure the overall quality of the benchmark execution as the Peak Signal to Noise Ratio (PSNR) between the input image and the image that is the outcome of a sequence of DCT, quantization, de-quantization and inverse-DCT operations. For Sobel, we measure the PSNR of the output of the unreliable execution with respect to an error-free execution. For all benchmarks, the error-free execution involves the scheduling of tasks on hardware which is configured to operate reliably.

The remainder of this section presents, quality and performance metrics for the ANN and the Topaz error detection methodologies for each benchmark. Performance overhead is calculated as the percentage of CPU cycles required for error detection and correction with respect to the cycles required to execute the application reliably. Note that, both error detection and correction are executed under reliable conditions.

\begin{figure}[tb]
\centering
\includegraphics[width=\linewidth]{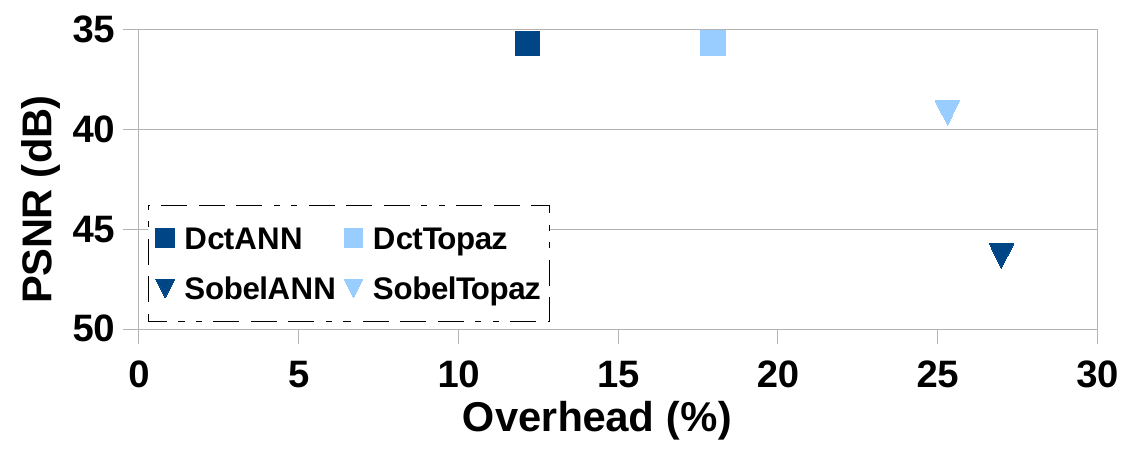}
\caption{DCT, Sobel Quality/Overhead}
\label{fig:qual_over_multimedia}
\includegraphics[width=\linewidth]{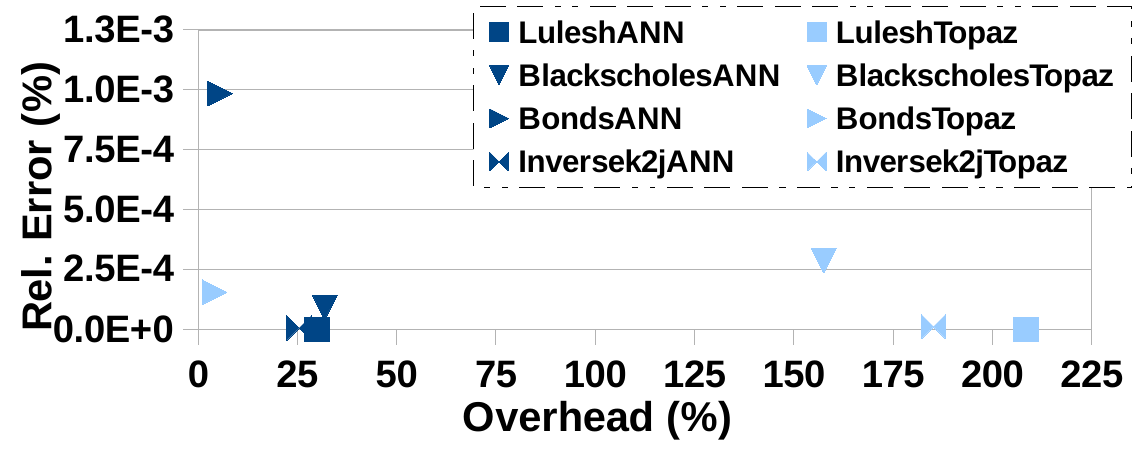}
\caption{Lulesh, Blackscholes, Bonds, Inversek2j Quality/Overhead}
\label{fig:qual_over_remaining}
\end{figure}

\begin{figure}[tb]
\centering
\includegraphics[width=\linewidth]{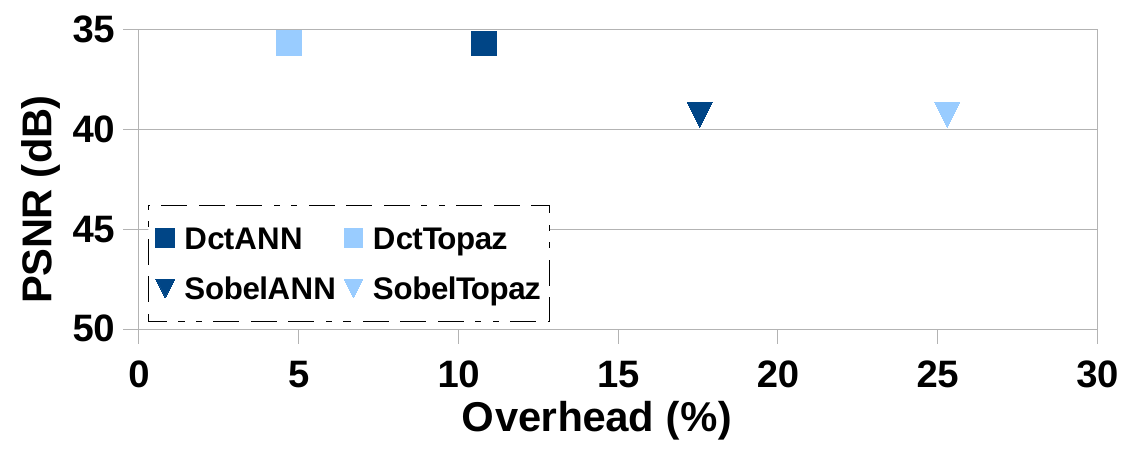}
\caption{DCT, Sobel Quality/Overhead (Batch)}
\label{fig:qual_over_batched_multimedia}

\includegraphics[width=\linewidth]{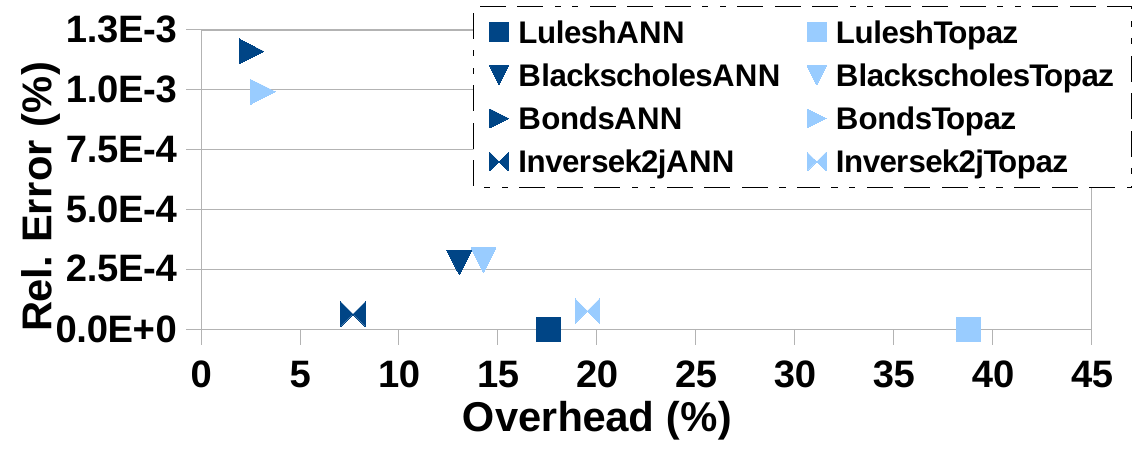}
\caption{Lulesh, Blackscholes, Bonds, Inversek2j Quality/Overhead (Batch)}
\label{fig:qual_over_batched_remaining}
\end{figure}

\begin{table}[tb]
\centering
\begin{tabular}{|l|cc|cc|}
\hline
&  \multicolumn{2}{c}{\textbf{Original}} & \multicolumn{2}{|c|}{\textbf{Batched}} \\
\cline{1-5}
\textbf{Benchmark} &  \textbf{ANN} & \textbf{Topaz} & \textbf{ANN} & \textbf{Topaz} \\
\hline
DCT &  \textbf{0.04} & 0.14                & 0.032           & \textbf{0.033}\\
\hline
Sobel & \textbf{3.63} & 3.82               & \textbf{1.26}   & 1.87\\
\hline
Lulesh & \textbf{6.1e-5} & $\infty$        & \textbf{3.1e-5} & $\infty$ \\
\hline
Blackscholes & \textbf{6.7e-4} & 1.7e-2    &1.3e-3           & \textbf{9.0e-4}\\
\hline
Bonds  & 1.7e-3 & \textbf{2.0e-4}          & \textbf{1.4e-3} & 1.5e-3 \\
\hline
Inversek2j & \textbf{6.6e-6} & $\infty$    & \textbf{1.6e-3} & 5.2e-3 \\
\hline
\end{tabular}
\caption{\metric{} values for all evaluated error-detectors. Lower numbers indicate more efficient detectors. Bold numbers indicate the fittest error detector. ANNs outperform Topaz in 5 and 4 out of 6 benchmarks when batching is toggled off and on respectively.}
\label{tbl:metric_comparison}
\end{table}

{\it Figures~\ref{fig:qual_over_multimedia}-\ref{fig:qual_over_batched_remaining}} present the Quality/Overhead measurements for both batch-enabled and non-batched versions of the $6$ benchmarks. For each benchmark we illustrate the results obtained using Topaz as well as the best performing ANN based on \metric{}. {\it Table~\ref{tbl:metric_comparison}} summarizes the \metric{} values.
On average, the per-benchmark fittest ANNs delivered an average \accuracyIncorrect{} of $94.85\%$ whereas Topaz scored marginally worse at $94.11\%$. Interestingly, the ANNs induce a much lower overhead in comparison to Topaz. In fact without the use of batched error-detection Topaz results in extreme overheads in 3 out of 6 benchmarks.

Both ANNs and Topaz are most efficient for applications with tasks that handle few data but execute for a long time. One example of such an application is Bonds. Tasks that perform few computations for large sets of data (i.e. Sobel) behave poorly with this type of error detection. Large feature vectors lead to an increased overhead for error detection.

Especially for multimedia applications we expect that there are better alternatives to Multilayer Perceptrons like Convolutional Neural Networks (CNNs)~\cite{lecun2015lenet}. CNNs have been designed to exploit the spatial information of data contained within a feature vector. This information is implicitly defined by the indices of the data within the vector. 

\subsection{Case study: An unreliable configuration at the PoFF}
\label{subsect:poff_analysis}

The Point Of First Failure (PoFF)~\cite{das2006self} denotes a point at which errors manifest at a frequency of one error every $10^7$ cycles. Hardware configurations which are closer to nominal values still exhibit errors, however at exponentially lower rates. Hardware operates at its PoFF when its voltage supply is approximately $85\%$ of the nominal value for the current frequency setting\cite{ReplayARM_JSSC2009}.  Every additional $10mV$ drop in voltage supply increases the fault rate exponentially by one order of magnitude\cite{das2006self}.

\begin{figure}[tb]
    \centering
       \centering
        \includegraphics[width=\columnwidth]{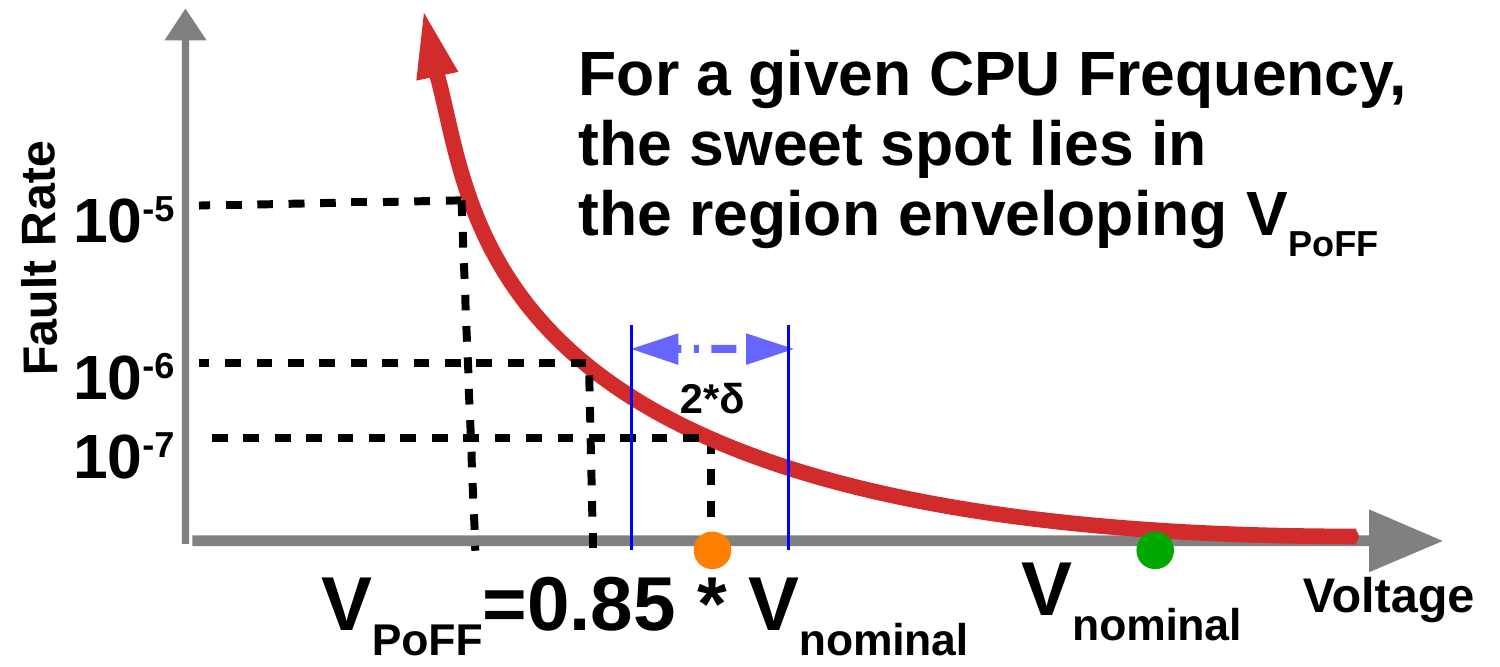}
        \caption{Fault rate, measured as a function of $\frac{1}{cycles}$, and CPU voltage for a specific CPU frequency}
        \label{fig:poff_rate_voltage}
\end{figure}

{\it Figure~\ref{fig:poff_rate_voltage}} shows the fault rate as a function of the supply voltage, when the clock frequency remains constant. There is a region ($\delta$) near $V_{PoFF}$ in which unreliable computing is the most efficient for improvements in performance, power, and/or energy~\cite{Parasyris:2017:SPE:3086564.3058980}. Moving closer to $V_{nominal}$ leads to less frequent faults at the cost of reduced performance/energy/power gains. But, the error detection overhead remains the same because all unreliably executed tasks need to be checked for errors, regardless of how likely it is for them to have to produced faulty results. Conversely, extreme sub-nominal voltage/frequency configurations, outside the nominal operating envelope, lead to higher fault rates and increase the correction overhead thereby outweighing any performance/energy/power benefits.

A CPU operates nominally (i.e. error free) on a (V, f) line. The CPU may dynamically move to multiple configurations between a lower and a higher performance point under the control of the Operating System (OS). For example, when the workload is low or memory-bound, the OS can switch the CPU core(s) closer to the lower performance point to save power. In this section, we evaluate whether we can reliably operate the CPU under unreliable configuration to reduce execution time using the proposed error detection technique.

For our experiments on the Intel i7 4820k CPU, we use two configurations. The first configuration is reliable ($V_{nominal}$, $f_{nominal}$) = ($0.9$ V, $1.67$ GHz). The second one is unreliable ($V_{nominal}$, $f_{overclocked}$) = ($0.9$ V, $3.7$ GHz). The unreliable configuration corresponds to the PoFF (note that $0.9 \approx 0.85*1.06$) and falls well into the unreliable area ({\it Figure~\ref{fig:poff_rate_voltage}}) due to overclocking. Overclocking provides lower execution time under unreliable conditions, which may result in crashes or SDCs. Error detection and correction mechanisms are then deployed to alleviate the effects of unreliable configurations, at the expense of performance (discussed in Section~\ref{subsect:analysis}). It is this interplay between output quality and performance (faster clock vs. correction/detection overhead) that we evaluate in this subsection.

We selected these specific (V, f) points to illustrate the performance gains of our methodology mainly because switching between them only requires a fast clock scaling~\cite{DynamicClockAdjustment_DATE2015}, rather than a slow voltage scaling mechanism. However, this analysis is applicable to any pair of reliable and unreliable operational points, including undervolting.
For our experiments which involve large input data sets we use the software fault injection methodology of Section \ref{sect:fault_injection} and inject faults at a rate of $1$ fault every $10^7$ cycles. To obtain an upper bound of the speedup, we introduce an oracle-like error detection mechanism with perfect \accuracyIncorrect{} (TPR=1, FNR=0) and zero detection overhead. The oracle mechanism still pays the overhead of error correction by task re-execution. 

{\it Table~\ref{tbl:oracle_comparison}} compares the fittest ANNs and Topaz against the Oracle for each benchmark.
The speedup baselines are error-free executions of the applications using the fully reliable voltage/frequency configuration of ($V_{nominal}$, $f_{nominal}$).
Keep in mind that for our applications we execute all tasks, unless otherwise specified, on unreliable cores which operate under the ($V_{nominal}$, $f_{overclocked}$) configuration. The remaining parts of the applications are executed under reliably ($V_{nominal}$, $f_{nominal}$).
The average speedup of the theoretical Oracle error detector is $1.99$x, the resulting output is by definition bit-wise exact since the Oracle always detects an \emph{Incorrect} task output. Even though both ANNs and Topaz exhibit similar behavior in terms of \accuracyIncorrect{}, the ANNs exhibit higher speed up at $1.51$ vs. $1.15$ for Topaz. The difference is mainly due to the higher error-detection overhead in Topaz (which is true also in the batched version of the benchmarks). 
The best scoring ANNs in batched configuration, on average, achieve an execution speedup of $1.72x$ which is larger than Topaz's $1.44$x. Again, the difference is mainly due to the high error detection overhead of Topaz for most benchmarks.

\begin{figure}[tb]
\includegraphics[width=0.95\columnwidth]{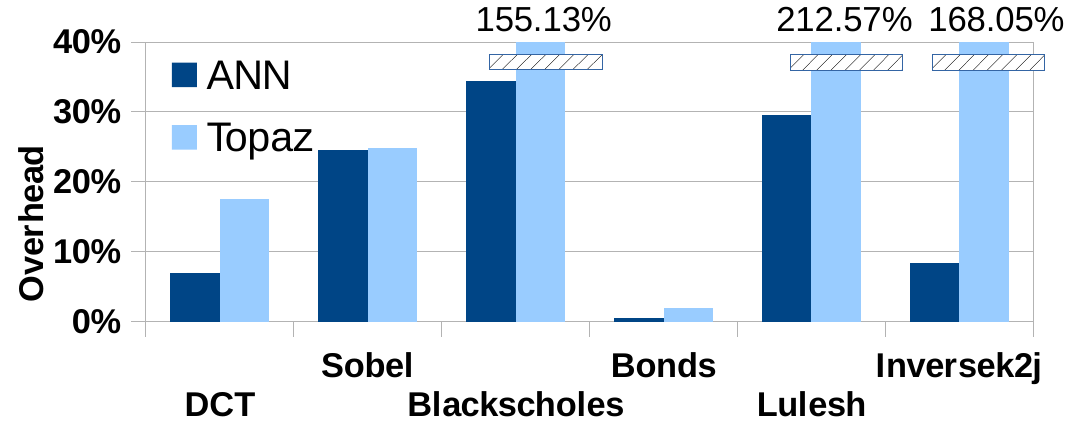}
\caption{The overhead of performing error detection defined as the number of cycles spent to perform error detection over the number of cycles required for a fully reliable execution of the application}
\label{fig:detection_overhead}
\end{figure}

\begin{figure}[tb]
\includegraphics[width=0.95\columnwidth]{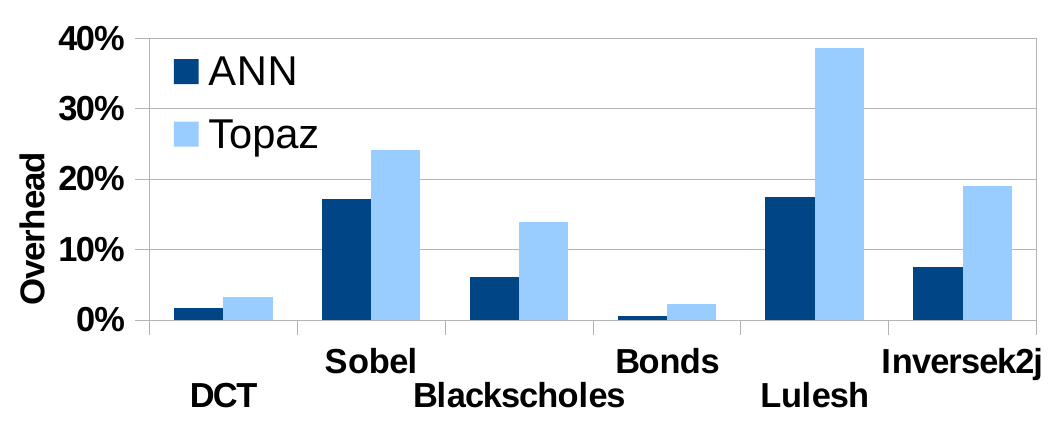}
\caption{Error detection and correction overhead for batched error-checking}
\label{fig:detection_overhead_batch}
\end{figure}

{\it Figure~\ref{fig:detection_overhead}} shows that the average error detection overhead with respect to the execution time of the application on a reliable CPU is $13.83\%$ and $72.98\%$ for the best ANN (the one with the lowest \metric{}), and Topaz, respectively. Note that the output quality levels are the same as the first case study. Three benchmarks (Blacksholes, Lulesh, inversej2k) have such a high detection overhead, that the whole application is slowing down as shown in {\it Table~\ref{tbl:oracle_comparison}}. For these benchmarks, Topaz is not a viable approach unless heavy task-batching is utilized to amortize the cost of error detection. A similar trend is observed across all benchmarks: performing error detection at a coarser level via batching reduces the average overhead of error detection. More specifically, the cost of error detection for ANNs after batching ({\it Figure~\ref{fig:detection_overhead_batch}}) comes to $6.45\%$ and for Topaz to $12.81\%$. Error-detection for applications which involve tasks with extremely small computational cost is only possible through batching. 

\begin{table}[tb]
\small
\centering
\setlength\tabcolsep{4.5pt}
\begin{tabular}{|l|c|cc|cc|}
\hline
&  & \multicolumn{2}{c}{\textbf{Original}} & \multicolumn{2}{|c|}{\textbf{Batched}} \\
\cline{2-6}
\textbf{Benchmark} & \textbf{Oracle} & \textbf{ANN} & \textbf{Topaz} & \textbf{ANN} & \textbf{Topaz} \\
\hline
DCT & 1.88 & 1.54 & 1.42          &1.51 &1.74\\
Sobel & 2.16 &1.36 & 1.40         &1.53 &1.40\\
Blackscholes & 2.11 & 1.17 & 0.51 &1.72 &1.62\\
Bonds  & 2.14 & 2.03 & 1.99       &2.10 &1.20\\
Lulesh &1.93 & 1.21 & 0.35        &1.48 &1.16\\
Inversek2j & 2.06 & 1.39 & 0.43   &1.85 & 1.41\\
\hline
Average & 1.99 & 1.51 & 1.15 & 1.72 & 1.44 \\
\hline
\end{tabular}
\caption{Comparison of speedups obtained from simulating overclocked hardware (($V_{nominal}$, $f_{overclocked}$) = ($0.9$ V, $3.7$ GHz)) vs the error-free baseline ($0.9$ V, $1.67$ GHz)}
\label{tbl:oracle_comparison}
\end{table}

%% file: 7_related_work.tex
\section{Related work}
\label{sect:7_related_work}

Current state of the art approaches to online error detection rely on duplicating the instructions of selected application parts which are considered error-prone. Unsafe instructions are first identified via compiler-analysis and/or profiling. Subsequently, a compiler pass hardens the application by duplicating the unsafe instructions and inserting checks~\cite{feng2010shoestring, lu2014sdctune, didehban2016nzdc, laguna2016ipas}. The checks typically involve redundancy in the form of instruction duplication. When a check detects an error the application is restored using some earlier checkpoint. IPAS~\cite{laguna2016ipas} expects the user to include a verification function that is used to check whether an injected fault has propagated to the output of the code which is targeted for software-hardening against soft-errors. This function is only used to train an ANN that drives the selection of instructions prior to their duplication. Other works \cite{hari2012low, grigorian2014dynamically,kadric2014energy} rely on manually implemented Light-Weight Checks (LWCs) to detect errors at the outputs of computations. In \cite{grigorian2014dynamically} LWCs are used to determine when an approximate alternative to a function computes outputs which severely differs from the exact implementation. \cite{kadric2014energy} relies on manually implemented LWCs to detect errors on the output of unreliably executed code. \cite{hari2012low} falls back to instruction duplication whenever light-weight error detectors result in low \accuracyIncorrect{}.

\cite{Ringenburg:2015:MDQ:2694344.2694365} presents two offline debugging mechanisms and three online monitoring mechanisms for approximate programs. The first offline mechanism identifies correlation between Quality of Result (QoR) and each approximate operation by tracking the execution and error frequencies of different code regions over multiple program executions with varying QoR values. The second mechanism tracks whether approximate operations affect some approximate variable and memory location. The online mechanisms complement the offline ones and they detect and compensate for QoR loss while maintaining the energy gains of approximation. The first one compares the QoR for precise and approximate variants of the program for a random subset of executions. This mechanism is useful for programs where QoR can be assessed by sampling a few outputs, but not for those that require bounding the worst-case errors. The second online mechanism uses programmer-supplied \emph{verification functions} that check a result with lower overhead than computing the result from scratch. The third mechanism stores past inputs and outputs of the checked code to estimate the output for current execution by interpolating past executions with similar inputs. It is shown that the offline mechanisms help in effectively identifying the root of a quality issue instead of merely confirming the existence of an issue, and that the online mechanisms help in controlling QoR while maintaining high energy gains. Our method may also detect errors due to approximation but in this work we focus on errors induced by unreliable execution.

\cite{Khudia:2015:ROQ:2872887.2750371} presents an output-quality monitoring and management technique which can ensure meeting a given output quality. Based on the observation that simple prediction approaches, (e.g. linear estimation, moving average, and decision trees) can accurately predict approximation errors, they use a low-overhead error detection module which tracks predicted errors to find the elements which need correction. Using this information, the recovery module, which runs in parallel to the detection module, re-executes the iterations that lead to high-errors. This becomes possible since the approximable functions or codes typically just simply read inputs and produce outputs without modifying any other state, such as map and stencil patterns. Large errors on the approximated computations are corrected by means of executing the accurate code using the CPU. Our approach differs in that we use an ANN to detect error whereas~\cite{Khudia:2015:ROQ:2872887.2750371} uses hardware accelerated ANNs to approximate code whose output is subsequently error checked. 

In one of the chronologically earlier efforts on task-based error-tolerant computing,\cite{Rinard_ICS2006} proposes a software mechanism that allows the programmer to identify task blocks and then creates a profile-driven probabilistic fault model for each task~. This is accomplished by injecting faults at task execution and observing the resulting output distortion and output failure rates. The concept of Task Level Vulnerability (TLV) captures dynamic circuit-level variability for each OpenMP task running in a specific processing core~\cite{omp_DATE2013}. TLV meta-data are gathered during execution by circuit sensors and error detection units to provide characterization at the context of an OpenMP task. Based on TLV meta-data, the OpenMP runtime apportions tasks to cores aiming at minimizing the number of instructions that incur errors. Unlike our work, they do not consider error recovery and user-specified approximate execution paths. 

An earlier work~\cite{Parasyris:2017:SPE:3086564.3058980} introduced an end-to-end framework for applications comprising mixed-criticality computation parts on heterogeneous mixed-reliability hardware. A programming model enables application developers specify computation in the form of tasks which are tagged with significance (criticality) information. At execution time, the most significant tasks are scheduled on reliable hardware, and the least ones on unreliable hardware. Unreliably executed tasks undergo a result-checking phase after their termination. Tasks which are considered to have produced erroneous output are then corrected either via re-execution or approximation. Error detection is achieved through the use of manually implemented error-checking functions.

%% file: 8_conclusions.tex
\section{Conclusions}
\label{sect:8_conclusions}

We presented a methodology for constructing, in a largely automated way, error-detectors that can identify, efficiently and effectively, erroneous outputs of selected parts of a program (tasks) to facilitate fault tolerant computing. We also provide a case study which exploits unreliable computing with CPU cores configured near their PoFF to gracefully trade-off output quality with performance. We employ light-weight error-detection through the use of low overhead ANNs. Those ANNs detect errors at the outputs of intermediate computations which would significantly impact the output quality of the application.
We showed that it is possible to rely on using ANNs for automatic result-checking which outperforms the well known approximate outlier detector Topaz~\cite{Topaz2015}. For most of the investigated applications an ANN incurs less computation overhead, and results in either better or equivalent output quality compared to Topaz. 
In fact, ANNs outperform Topaz for real unreliable execution for all the benchmarks that we tried on overclocked ARM Cortex A53 CPUs.

Moreover, we observed that batching can significantly reduce the overhead of error detection. Unfortunately, that comes with a cost to the \errorAccepted{}. This increased \errorAccepted{} does not severely affect the overall output quality of applications because the \accuracyIncorrect{} of the error detectors tends to remain very high even with batching. Our methodology enables a number of interesting future research directions for designing dynamic runtime systems. For example, one can automatically generate a number of error detectors which vary in terms of overhead and resulting output quality. At execution time, the runtime system can choose the best error detector depending on their \metric{}, \errorAccepted{} scores, and user supplied constraints such as a maximum energy budget for error detection. Finally, our methodology generates small and lightweight ANNs which in turns motivates the investigation of policies regarding fine-tuning during execution time.